\documentclass[journal]{IEEEtran}
\usepackage{amsmath}
\usepackage{amsfonts}
\usepackage{bm}
\usepackage{eqlist}
\usepackage{subfigure}
\usepackage{graphicx}
\usepackage{epstopdf}
\usepackage{color}
\usepackage{placeins}
\usepackage{float}
\usepackage{cite}
\usepackage{multirow}
\usepackage{tabularx,colortbl}
\usepackage{siunitx}
\usepackage{pgfplots}
\pgfplotsset{compat=1.11, legend image code/.code={
				\draw[mark repeat=2,mark phase=2] plot coordinates {
				(0cm,0cm)
				(0.4cm,0cm)        %% default is (0.3cm,0cm)
				(0.8cm,0cm)         %% default is (0.6cm,0cm)
				};%
			}
    }
\usepgfplotslibrary{polar}
\usetikzlibrary{spy}
\usetikzlibrary{external}
\tikzset{external/export=true}
\tikzset{external/force remake=false}

\newlength\figureheight
\newlength\figurewidth

\usepackage{hyperref}

\newcommand{\figref}{Fig.~\ref}

\newcommand{\secref}{Sec.~\ref}

\newcommand{\mb}{\textrm{mb}}

\newcommand{\ga}{\textrm{ga}}
\newcommand{\bloch}{\textrm{B}}
\newcommand{\even}{\textrm{e}}
\newcommand{\odd}{\textrm{o}}

\newcommand{\temp}{\textrm{t}}
\newcommand{\spat}{\textrm{s}}
\newcommand{\ep}{\textrm{EP}}

\newcommand{\pati}[1]{}

\begin{document}

\title{Theory of Seamless-Scanning \\ Periodic Leaky-Wave Antennas based \\ on $\mathcal{PT}$-Symmetry with Time to Space Mapping}
\author{Amar~Al-Bassam,~\IEEEmembership{Member,~IEEE,}
        Simon~Otto,~\IEEEmembership{Member,~IEEE,}
		Dirk~Heberling,~\IEEEmembership{Senior~Member,~IEEE,}
        and~Christophe~Caloz,~\IEEEmembership{Fellow,~IEEE}
}
%\thanks{A.~Al-Bassam and D.~Heberling are with the Institute of High Frequency Technology, RWTH Aachen University, 52074 Aachen, Germany.}
%\thanks{D.~Heberling is also with the Fraunhofer Institute for High Frequency Physics FHR, 53345 Wachtberg, Germany.}
%\thanks{S.~Otto is with IMST GmbH, 47475 Kamp-Lintfort, Germany.}
%\thanks{C.~Caloz is with ESAT-WaveCore-META, KU Leuven, 3001 Leuven, Belgium}
%\thanks{Manuscript received month xx, xxxx; revised month xx, xxxx.}}
%\markboth{IEEE TRANSACTIONS ON ANTENNAS AND PROPAGATION, VOL. x, NO. xx, xxxx}
%{Shell \MakeLowercase{\textit{et al.}}: Bare Demo of IEEEtran.cls for Journals}

\maketitle

%\tableofcontents

%%%%%%%%%%%%%%%%%%%%%%%%%%%%%%%%%%%%%%%%%%%%%%%%%%%%%%%%%%%%%%%%%%%%%%%%%%%%%%%%%%%%%%%
%%%%%%%%%%%%%%%%%%%%%%%%%%%%      Abstract    %%%%%%%%%%%%%%%%%%%%%%%%%%%%%%%%%%%%%%%%%
%%%%%%%%%%%%%%%%%%%%%%%%%%%%%%%%%%%%%%%%%%%%%%%%%%%%%%%%%%%%%%%%%%%%%%%%%%%%%%%%%%%%%%%
\begin{abstract}
Periodic Leaky-Wave Antennas (P-LWA) offer highly directive and space-scanning radiation. Unfortunately, they have been plagued by the ``broadside issue'', characterized by a degradation in gain when the antenna's main beam is steered across broadside. While this issue has been addressed by circuit and network approaches, a related fundamental and general electromagnetic theory has been lacking. This paper fills this gap. We first show that a P-LWA is a $\mathcal{PT}$-symmetric system, whose even- and odd-mode coupling in the complex space of temporal frequencies leads to the characteristic pair of two-sheet Riemann surfaces. We observe that the branch cuts of these surfaces, which form the well-known $\mathcal{PT}$-symmetric double pitchfork spectrum, correspond to the ``balanced frequency'' condition, while the branch point at the junction of the pitchforks, is an exceptional point that corresponds to the ``$Q$-balanced'' condition, two conditions that where previously shown to be the conditions for eliminating the broadside issue. In order to acquire an independent and rigorous interpretation of this spectrum, we further transform the coupled complex temporal eigenfrequencies into complex spatial eigenfrequencies. We identify the resulting spatial frequencies as coupled forward-backward modes, with frequency-independent imaginary parts (leakage factors), a condition for P-LWA equalization across broadside. Finally, we derive the scattering parameters of the P-LWA and show that matching is achieved only at one of the two ends of the P-LWA structure. This work both provides a solid foundation to the theory of P-LWAs and represents an original contribution to the field of $\mathcal{PT}$-symmetry.
\end{abstract}


\IEEEpeerreviewmaketitle

%%%%%%%%%%%%%%%%%%%%%%%%%%%%%%%%%%%%%
%%%%%%%%% List of Symbols  %%%%%%%%%%
%%%%%%%%%%%%%%%%%%%%%%%%%%%%%%%%%%%%%
\section{List of Symbols}\label{sec:los}
\vspace{2mm}
\begin{tabularx}{\linewidth}{@{}>{}l@{\hspace{.6em}}X@{}}
\multicolumn{2}{l}{\sc{Time Analysis (Resonator Problem)}}\\
$x_{\even,\odd}$    & even- (or symmetric-) mode or odd- (or antisymmetric-) mode eigenquantity $x$ \\
$\psi_{\temp;\even,\odd}$ & complex temporal eigenfrequency electromagnetic field:  $\psi_{\temp;\even,\odd}(t,z)=\exp(j\xi_{\even,\odd}{t})f_{\even,\odd}(\beta{z})$, \\
$\xi$               & complex temporal eigenfrequency: $\xi=\omega+j\zeta$ \\
$\omega$            & real part of $\xi$, i.e., usual (angular) frequency \\
$\zeta$             & imaginary part of $\xi$, i.e., inverse of the relaxation time \\
$Q$                 & quality factor: $Q=\omega/(2\zeta)$  \\ 
$\Phi$              & periodic boundary condition (PBC) phase shift, i.e., phase shift between the two terminals of the unit cell \\
$x_{\temp;{1,2}}$   & coupled-mode (coupling between the even and odd eigenmodes) temporal eigenquantity $x_\temp$ number 1 or 2 \\
$\psi_{\temp;1,2}$  & coupled-mode electromagnetic field: \\
                    & $\psi_{\temp;1,2}(t,z)=\exp(j\xi_{1,2}{t})f_{1,2}(\beta{z})$, where $f_{1,2}(\cdot)=a_{1,2}\cos(\cdot)+b_{1,2}\sin(\cdot)$ \\
$\kappa_\temp$      & temporal coupling factor between even and odd modes \\
$\kappa_{\temp,\ga}$& temporal coupling factor due to geometrical asymmetry \\
$\kappa_\ep$        & temporal coupling factor at the exceptional point
\end{tabularx}

\vspace{3mm}

\begin{tabularx}{\linewidth}{@{}>{}l@{\hspace{.6em}}X@{}}
\multicolumn{2}{l}{\sc{Space Analysis (Waveguide Problem)}}\\
$x_\pm$             & forward ($+$) and backward ($-$) space harmonic (SH) \\
$\psi_{\spat,\pm}$  & complex spatial eigenfrequency electromagnetic field: $\psi_{\spat,\pm}(z,t)=\exp(-\gamma_\pm{z})\exp(j\omega{t})$ \\ 
$\gamma$            & complex spatial frequency, or wavenumber: $\gamma=\alpha+j\beta$ \\
$\alpha$            & real part of $\gamma$, or attenuation constant \\
$\beta$             & imaginary part of $\gamma$, or phase constant \\
$x_{\spat;{1,2}}$   & coupled-mode (between the forward and backward SHs) eigenquantity $x_\spat$ number 1 or 2 \\
$\psi_{\spat;1,2}$  & spatial coupled-mode electromagnetic field:\\
                    & $\psi_{\spat;1,2}(t,z)=\exp(-\gamma_{1,2}{z})\exp(j\omega{t})$ \\
$\kappa_\spat$      & spatial coupling factor between the forward and backward SHs \\
$\kappa_{\spat,\ga}$& spatial coupling factor due to geometric asymmetry \\
$\theta_\mb$        & radiation angle of the main beam of the P-LWA: $\theta_\mb \approx \sin^{-1}(\beta/k_0)$, where $k_0$ is the free-space wavenumber, $k_0=\omega/c$, with $c$ being the speed of light in free-space
\end{tabularx}

%%%%%%%%%%%%%%%%%%%%%%%%%%%%%%%%%%%%%
%%%%%%%%%%% Introduction  %%%%%%%%%%%
%%%%%%%%%%%%%%%%%%%%%%%%%%%%%%%%%%%%%
\section{Introduction}\label{sec:intro}

\begin{figure}
    \centering
    \includegraphics[width=0.5\textwidth]{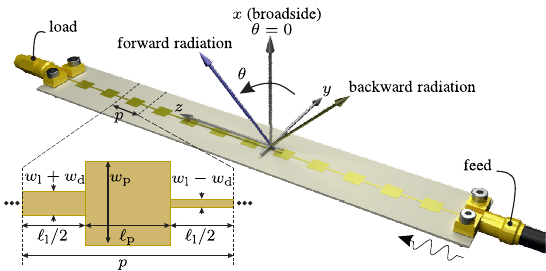}
    \caption{Structure and operation of a Periodic Leaky-Wave Antenna (P-LWA), illustrated through the specific case of a Series-Fed Patch (SFP) design. This SFP-P-LWA~\cite{Al-Bassam_APS_07_2023} will be used as a representative example throughout the paper; however, the proposed theory is applicable to any P-LWA. The geometrical parameters  $\ell_\textrm{l} = \SI{3.5}{mm}$, $\ell_\textrm{p} = \SI{3.05}{mm}$, $w_\textrm{l} = \SI{0.3}{mm}$ and $w_\textrm{p} = \SI{3.3}{mm}$ with a substrate height of $\SI{508}{\micro m}$ and $\epsilon_r = 3.66$ are kept constant throughout the paper, while the parameter $w_\textrm{d}$ is varied, with $w_\textrm{d}=0$ and $w_\textrm{d}\neq{0}$ representing a symmetric and an asymmetric structure, respectively, with respect to the transverse ($y$) direction.}
    \label{fig:P-LWA}
\end{figure}

\pati{Generalities on Leaky-Wave Antennas}
Leaky-wave antennas are traveling-wave structures that support faster-than-light waves, enabling electromagnetic radiation with frequency scanning~\cite{Jackson_Balanis_book_chap,Caloz_McGrawHill_Gross_Book_2011,Jackson_PIEEE_07_2012}. These antennas can be uniform or periodic; however, the former are limited to radiating within a partial sector of space~\cite{Menzel_EuMC_10_1978}, while the latter provide full-space scanning capability~\cite{Jackson_PIEEE_07_2012}. Therefore, this paper will focus on Periodic Leaky-Wave Antennas (P-LWAs), illustrated by the example in \figref{fig:P-LWA}. Being non-resonant, such antennas can be electrically very long, providing significant directivity, without requiring a complex feeding network as antenna arrays. Consequently, they find applications in many areas, including communication~\cite{Abielmona_TAP_04_2011}, real-time spectral analysis~\cite{Gupta_MTT_12_2009}, sensing~\cite{Caekenberghe_RADAR_04_2007} and imaging~\cite{Li_TTHZ_05_2016}.

\begin{figure}
    \centering
    \setlength\figureheight{.2\textwidth}
    \setlength\figurewidth{.2\textwidth}
    % \mbox{
    %     \tikzsetnextfilename{BSP_SB}
    %     \subfigure[]{\input{figs/SFP_Gain_vs_theta_Sym.tikz}\label{fig:rad_pat_bs1}}
    %     \tikzsetnextfilename{BSP_DB}
    %     \subfigure[]{\input{figs/SFP_Gain_vs_theta_Asym_P1.tikz}\label{fig:rad_pat_bs2}}
    % }
    %\mbox{
    \subfigure[]{\includegraphics[]{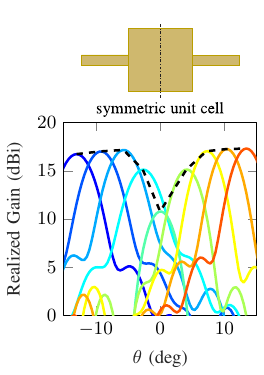}\label{fig:rad_pat_bs1}}
    \subfigure[]{\includegraphics[]{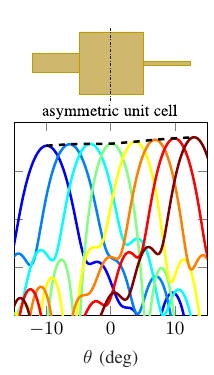}\label{fig:rad_pat_bs2}}
    %}
    \caption{Broadside radiation patterns at different frequencies around broadside for the SFP-P-LWA in \figref{fig:P-LWA} with (a)~a symmetric unit cell and (b)~an optimally asymmetric unit cell.}
    \label{fig:rad_pat_bs}
\end{figure}

\pati{The Broadside Issue and its Resolution}
Since their inception over eighty years ago~\cite{Goldstone_TAP_10_1959}, P-LWAs have been plagued by the so-called \emph{``broadside issue''}~\cite{Jackson_Balanis_book_chap, Caloz_McGrawHill_Gross_Book_2011}, characterized by a degradation in gain when the antenna's main beam is steered across broadside ($\theta=0$ in \figref{fig:P-LWA}). This phenomenon occurs due to the phase matching of forward and backward space-harmonics, resulting in a standing wave regime that creates an ``open stop-band''~\cite{Jackson_Balanis_book_chap},\footnote{The term ``open'' in ``open stopband'' refers to the fact that the stopband occurs in an open (radiative) system rather than in a closed (wave-guided)~one.} reflecting most of this incident energy back to the source. The broadside issue was addressed in recent years, using a transmission-line network matching approach~\cite{Paulotto_TAP_07_2009} and a circuit modeling approach~\cite{Otto_TAP_10_2011, Otto_TAP_04_2014, Otto_TAP_10_2014, Al-Bassam_TAP_06_2017}, both revealing the necessity of structural \emph{asymmetry}. Figure~\ref{fig:rad_pat_bs} presents typical radiation patterns for a P-LWA, with \figref{fig:rad_pat_bs}(a) showing the broadside gain degradation for a symmetric structure (the broadside issue)\footnote{The level of gain degradation at broadside varies among structures. Specifically, the structure in the example of \figref{fig:P-LWA} still considerably radiates at broadside, about $\SI{6}{dB}$ below the off-broadside gain (\figref{fig:rad_pat_bs}), due to relatively large transverse ($x$-direction) radiative metallic parts (here patches).} and \figref{fig:rad_pat_bs}(b) showing the broadside gain equalization for a structure with optimal asymmetry (resolution of the broadside issue). However, a \emph{comprehensive theory} rigorously explaining these symmetry-related phenomena has remained elusive to date.

\pati{Contribution: $\mathcal{PT}$-Symmetry Theory}
Here, we address this gap by presenting the first complete and rigorous electromagnetic theory of the broadside issue, and, by extension, of P-LWAs in general, using the powerful concept of parity-time symmetry, or $\mathcal{PT}$-symmetry~\cite{Bender_PRL_06_1998, Bender_PT_Book_2019}. Specifically, we demonstrate that a P-LWA is a $\mathcal{PT}$-symmetric system with an Exceptional Point (EP), and show that operating the antenna at this EP is the key to achieving equalized gain across broadside or, equivalently, seamless full-space scanning.

\pati{Organization of the Paper}
The paper is organized as follows. Following the list of symbols in \secref{sec:los} and this introduction, \secref{sec:intro}, \secref{sec:PT} recalls the fundamentals of $\mathcal{PT}$-symmetry, including the underpinning notion of gain-loss systems, the mathematical formulation of $\mathcal{PT}$-symmetry in terms of commutation, the related complex eigenspectra and their EP. Then, \secref{sec:TA_RP} conducts a time-domain analysis of the P-LWA system that determines its coupled-mode complex eigenfrequencies, plots the corresponding Riemann surfaces and points out the related $\mathcal{PT}$-symmetry, identifies its EP in the frequency balanced regime, and shows that this EP corresponds to the $Q$-balancing condition described in~\cite{Otto_TAP_10_2014}. Next, \secref{sec:TS_mapping} maps these complex temporal eigenfrequencies to their spatial counterparts, demonstrating that the EP corresponds to the real part of the spatial eigenfrequency (or leakage factor) being independent of frequency, a condition that enables uniform scanning through broadside. Finally, \secref{sec:SA_WP} performs a space analysis that determines the related coupled complex eigenfrequencies and derives the scattering coefficients in the forward and backward directions, revealing that one direction is matched while the other is not. Concluding remarks are presented in \secref{sec:concl}.

% %%%%%%%%%%%%%%%%%%%%%%%%%%%%%%%%%%%%%
% %%%%%%%% PT-Symmetry and EPs %%%%%%%%
% %%%%%%%%%%%%%%%%%%%%%%%%%%%%%%%%%%%%%
\section{\texorpdfstring{$\mathcal{PT}$}{PT}-Symmetry and Exceptional Points}\label{sec:PT}

\begin{figure}
    \centering
    \subfigure[]{\includegraphics{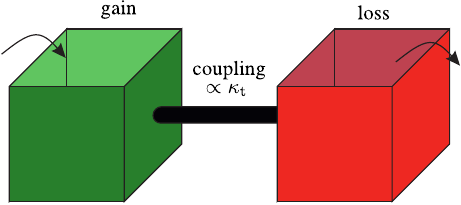}\label{fig:PT_pipe}}
    \subfigure[]{\includegraphics{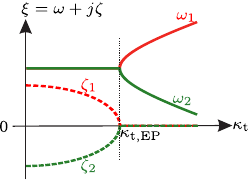}\label{fig:PT_gg}}
    \subfigure[]{\includegraphics{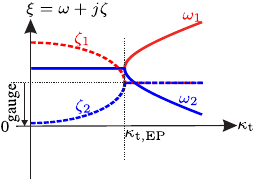}\label{fig:PT_gl}}
    \caption{$\mathcal{PT}$-symmetric system, composed of a gain box (box into which energy is added) and a loss box (box from which energy is removed), coupled by a duct (adapted from from~\cite{Bender_PT_Book_2019}). (a)~Coupled-resonator system. (b)~Complex eigenfrequencies of the system in~(a) versus its coupling coefficient, $\kappa_\temp$. (c)~Complex eigenfrequencies of the gauge-transformed loss-loss version of the system in~(a).}
    \label{fig:PTS_gauge}
\end{figure}

\pati{$\mathcal{PT}$ Origin and Definition}
$\mathcal{PT}$-symmetry is a concept that originates in 
quantum mechanics~\cite{Bender_PRL_06_1998}, but that also applies to classical optics~\cite{Hodaei_Science_11_2014,Peng_NPHY_04_2014,Zhang_SR_04_2016,Feng_NPHO_11_2017,Oezdemir_NMAT_08_2019,Gupta_AM_12_2019}. A system is said to be $\mathcal{PT}$-symmetric if its non-Hermitian and yet has real eigenvalues. Such a situation occurs in \emph{coupled gain-loss systems}~\cite{Bender_PT_Book_2019, Oezdemir_NMAT_08_2019}, such as the system
shown in \figref{fig:PTS_gauge}(a), featuring two interconnected box resonators—-one with gain and the other with loss—-linked by a coupling duct. This system exhibits $\mathcal{PT}$-symmetry because its evolution is identical if i)~it is successively subjected to a $\mathcal{T}$ or time reversal operations ($t\rightarrow-t$), equivalent to swapping gain and loss, and a $\mathcal{P}$ or space reversal operation ($x\rightarrow-x$), equivalent to exchanging the two boxes, and ii)~if the time reversal and space reversal operations are performed in the opposite order.

\pati{$\mathcal{PT}$ Mathematical Statement}
Mathematically, the $\mathcal{PT}$-symmetry of a system described by the Hamiltonian $\mathcal{H}$ may be formulated as~\cite{Bender_PT_Book_2019}
\begin{subequations}\label{eq:PTS_def}
    \begin{equation}\label{eq:PTS_def_a}
        \begin{split}
            (\mathcal{PT})\mathcal{H}(\mathcal{PT})^{-1}
            &=(\mathcal{PT})\mathcal{H}(\mathcal{T}^{-1}\mathcal{P}^{-1}) \\
            &=\mathcal{P}(\mathcal{T}\mathcal{H}\mathcal{T}^{-1})\mathcal{P}^{-1} \\
            &=\mathcal{P}\mathcal{H}^*\mathcal{P} \\
            &=\mathcal{H},
        \end{split}
    \end{equation}
    or, equivalently, by the commutation
    \begin{equation}
        \left[\mathcal{PT},\mathcal{H}\right]
        =(\mathcal{PT})\mathcal{H}-\mathcal{H}(\mathcal{PT})=0.
    \end{equation}
\end{subequations}
In the second equality of Eq.~\eqref{eq:PTS_def_a}, $\mathcal{THT}^{-1}$ represents the effect of time reversal on $\mathcal{H}$, which amounts to taking the conjugate of $\mathcal{H}$ (i.e., replacing its $ji$'s by $-j$'s), while in the third equality of Eq.~\eqref{eq:PTS_def_a}, $\mathcal{P}\mathcal{H}^*\mathcal{P}$ represents the effect of space reversal on $\mathcal{H}^*$, where $\mathcal{P}$ is the (spatial) exchange operator $[0,1;1,0]$~\cite{Bender_PT_Book_2019}
%\footnote{If the boxes are exactly identical and their gain and loss exactly compensate, then the $\mathcal{PT}$ operation leaves the system in its original state, i.e., $(\mathcal{PT})\mathcal{H}=\mathcal{H}$, or $\mathcal{PT}=\mathcal{I}$, where $\mathcal{I}$ is the identity operator. However, this is a very particular case of $\mathcal{PT}$-symmetry that represents an excessively strong constraint. In general, if the boxes are different or the gain and loss do not compensate, $(\mathcal{PT})\mathcal{H}\neq\mathcal{H}$, and only the condition in Eq.~\eqref{eq:PTS_def}, which is really the definition of $\mathcal{PT}$ symmetry, is aP-LWAys satisfied.}
.

\pati{Exceptional Points}
$\mathcal{PT}$-symmetric systems exhibit a pair of complex eigenfrequencies, $\xi_{1,2}=\omega_{1,2}+j\zeta_{1,2}$, as illustrated in \figref{fig:PT_gg}~\cite{Bender_PT_Book_2019}. These eigenfrequencies correspond to two double-sheet Riemann surfaces, as will be shown in the next section. Assuming the time harmonic dependence $\exp(j\xi{t})$, $\zeta_1$ and $\zeta_2$ correspond to loss and gain, respectively. These values merge to the value zero at the junction of the two pitchforks, at a specific point of the coupling parameter space, $\kappa_\temp$. This point corresponds to an exact balance between gain and loss in the system, and is called \emph{Exceptional Point} (EP).

\pati{Passive $\mathcal{PT}$-Symmetry}
A $\mathcal{PT}$-symmetric system may also be purely passive. This occurs when the imaginary part of the eigenspectrum in \figref{fig:PT_gg} is shifted upwards, transforming the negative (gain) $\zeta_2$ into a purely positive (loss) value, while the real part of the eigenspectrum remains unchanged, as shown in \figref{fig:PT_gl}. Mathematically, this shift corresponds to a \emph{gauge transformation}, which will be explained in the following section.

% %%%%%%%%%%%%%%%%%%%%%%%%%%%%%%%%%%%%%
% %%% Temporal Analysis  %%%%
% %%%%%%%%%%%%%%%%%%%%%%%%%%%%%%%%%%%%%
\section{Time Analysis (Resonator Problem)}\label{sec:TA_RP}
\pati{P-LWA relation with a $\mathcal{PT}$-symmetry system}
A priori, a P-LWA (\figref{fig:P-LWA}) seems quite different from a $\mathcal{PT}$-symmetry system (\figref{fig:PT_pipe}). However, it is closely related to such a system because its unit cell also supports two coupled resonant modes, or eigenmodes, and consequently also exhibits a complex eigenspectrum of the type shown in \figref{fig:PT_gg} or \figref{fig:PT_gl}. In order to apply $\mathcal{PT}$-symmetry theory, we first consider, in this section, the \emph{temporal frequency} domain, where the unit cell is excited as a resonator and exhibits therefore complex eigenfrequencies. This domain is the space where one may apply periodic boundary conditions along the (irreducible) Brillouin zone to compute the dispersion diagram of the structure\cite{Caloz_MetaTra_2006}. However, the following analysis will focus on the $\Gamma$ spectral point, i.e., $\Phi=\beta{p}=0$, where $\Phi$ is the phase shift across the unit cell, $\beta$ is the wavenumber and $p$ is the period of the structure; this point, which also implies $\beta=0$, corresponds to the broadside radiation regime~\cite{Caloz_MetaTra_2006}.

\begin{figure}
    \centering
	\mbox{
    \subfigure[]{\includegraphics[width=0.23\textwidth]{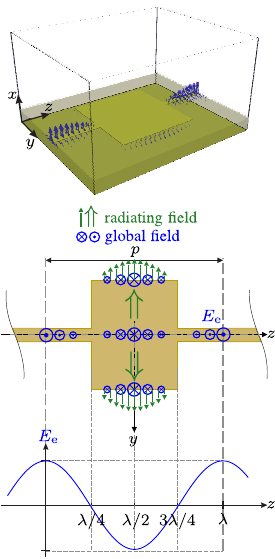}\label{fig:SFP_EVEN}}
    \subfigure[]{\includegraphics[width=0.23\textwidth]{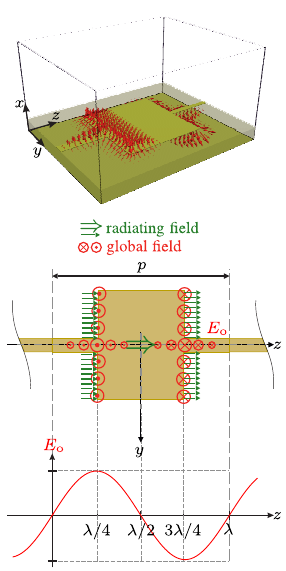}\label{fig:SFP_ODD}}
    }
    \caption{Complex eigenmodes for the SFP-P-LWA in \figref{fig:P-LWA} or \figref{fig:rad_pat_bs}. (a)~Even (or symmetric) eigenmode, $\psi_{\temp,\even}$, with eigenfrequency $\xi_{\even}=\omega_{\even}+j\zeta_{\even}$. (b)~Odd (or asymmetric) eigenmode, $\psi_{\temp,\odd}$, with eigenfrequency $\xi_{\odd}=\omega_{\odd}+j\zeta_{\odd}$.}
    \label{fig:eigen_modes}
\end{figure}

\pati{P-LWA Even and Odd Modes}
The two P-LWA modes are shown in \figref{fig:eigen_modes}.\footnote{In contrast to the two-box system in \figref{fig:PT_pipe}, the P-LWA has eigenmode that are essentially \emph{collocated}, i.e., present at the same location of space, but this does not pose any problem as long as their coupling can also be modeled by a coupling coefficient, as will be the case here.} One of them, depicted in \figref{fig:SFP_EVEN}, is a symmetric or \emph{even} mode that is obtained by terminating the unit cell at its two ends with a Perfect Magnetic Conductor (PMC) wall or, equivalently, an open circuit. The other mode, depicted in \figref{fig:SFP_ODD}, is an antisymmetric or \emph{odd} mode that is obtained by terminating the unit cell at its two ends with a Perfect Electric Conductor (PEC) wall or, equivalently, a short circuit. The two eigenmodes in \figref{fig:eigen_modes} may be written as\footnote{The t subscript in the following expressions emphasizes the related quantities pertains to the domain of temporal complex frequencies, or resonator regime.}$^\textrm{,}$\footnote{We omit here the transverse variables, $x$ and $y$, as the related dependencies do not play a role in the analysis. Accordingly, the functions $\psi$ represent the part of the (electric or magnetic) field that depends only on the spatial variable~$z$.}
\begin{subequations}\label{eq:cpl_mode_ansatz}
    \begin{equation}\label{eq:even_mode_psi}
        \psi_{\temp,\even}=\exp\left(j\xi_{\even}t\right)f_\even\left(z\right) 
    \end{equation}
    and
    \begin{equation}\label{eq:odd_mode_psi}
        \psi_{\temp,\odd}=\exp\left(j\xi_{\odd}t\right)f_\odd\left(z\right),
    \end{equation}
\end{subequations}
where the temporal and spatial dependencies have been assumed to be independent of each other, with $f_\even(z)$ and $f_\odd(z)$ being even and odd functions of space along the direction of the antenna axis ($z$ direction), and where
\begin{subequations}\label{cplx_eo_freq}
    \begin{equation}
        \xi_{\even}=\omega_{\even}+j\zeta_{\even},
    \end{equation}
    and
    \begin{equation}
        \xi_{\odd}=\omega_{\odd}+j\zeta_{\odd},
    \end{equation}    
\end{subequations}
are the even and odd complex frequencies, where the real parts, $\omega_{\even}$ and $\omega_{\odd}$, represent the resonators' oscillation frequencies, while the imaginary parts, $\zeta_{\even}$ and $\zeta_{\odd}$, represent the related power decay once the source has been removed. 

\pati{Coupled-Mode Equations}
In the absence of coupling between the even and odd modes, the P-LWA eigenstates are described by the uncoupled equations
\begin{subequations}\label{eq:tcmt}
 \begin{align}   
  \begin{split}     
    \dfrac{d{\psi_{\temp,\even}}}{dt} = j\xi_{\even} \psi_{\temp,\even},
  \end{split}\\
  \begin{split}    
    \dfrac{d{\psi_{\temp,\odd}}}{dt} = j\xi_{\odd} \psi_{\temp,\odd},
  \end{split} 
 \end{align}
\end{subequations}
which directly follow from the ans\"{a}tze in Eq.~\eqref{eq:cpl_mode_ansatz}. In contrast, in the presence of coupling, the governing equations of the system become the coupled-mode equations
\begin{subequations}\label{eq:cme}
 \begin{align}   
  \begin{split}     
   \dfrac{d{\psi_{\temp,1}}}{dt} = j\xi_\even{\psi_{\temp,1}} + j\kappa_{\temp,\even\odd}{\psi_{\temp,2}},   
  \end{split}\\
  \begin{split}    
   \dfrac{d{\psi_{\temp,2}}}{dt} = j\kappa_{\temp,\odd\even}{\psi_{\temp,1}} + j\xi_\odd{\psi_{\temp,2}},
  \end{split} 
 \end{align}  
 \label{eq:coupledEquation} 
\end{subequations}
where $\psi_{\temp,1}$ and $\psi_{\temp,2}$ are the new modes resulting from the coupling of the even and odd modes, which may be written
\begin{subequations}
    \begin{equation}\label{eq:cpl_mode_psi}
        \psi_{\temp;1,2}(t)=\exp(j\xi_{1,2}t),
    \end{equation}   
    with
    \begin{equation}\label{eq:cpl_mode_xi}
        \xi_{1,2}=\omega_{1,2}+j\zeta_{1,2},
    \end{equation}   
\end{subequations}
and where $\kappa_{\temp,\odd\even}$ and $\kappa_{\temp,\even\odd}$ are the coupling coefficients from the even mode to the odd mode and vice-versa, which are related as 
\begin{align}\label{eq:kappa_oe}
    \kappa_{\temp,\even\odd}=\kappa_{\temp,\odd\even}^*=\kappa_\temp,
\end{align}
by reciprocity~\cite{Caloz_PR_10_2018}.\footnote{Specifically, this equality is justified as follows. The factors $\kappa_{\temp,\even\odd}$ and $\kappa_{\temp,\odd\even}$ represent exchanges of energy in opposite directions between the eigenmodes, and changing direction corresponds to time reversal or, equivalently, phase conjugation; so, $\kappa_{\temp, \odd\even}^*$ corresponds to a reversal of exchange direction with respect to $\kappa_{\temp,\even\odd}$. Moreover, the system has no external force (biasing field or modulation); therefore, the system must be time-reversal symmetric. Thus, one must have $\kappa_{\temp,\even\odd}=\kappa_{\temp,\odd\even}^*$.}

\begin{figure}
    \centering
    \setlength\figureheight{.25\textwidth}
    \setlength\figurewidth{.25\textwidth}
    % \tikzsetnextfilename{RiemannSheet_Re}
    %\subfigure[]{\input{figs/RiemannSheet_Re2.tikz}\label{fig:Riemann_Re}}
    % \tikzsetnextfilename{RiemannSheet_Im}
    %\subfigure[]{\input{figs/RiemannSheet_Im2.tikz}\label{fig:Riemann_Im}}
    \subfigure[]{\includegraphics{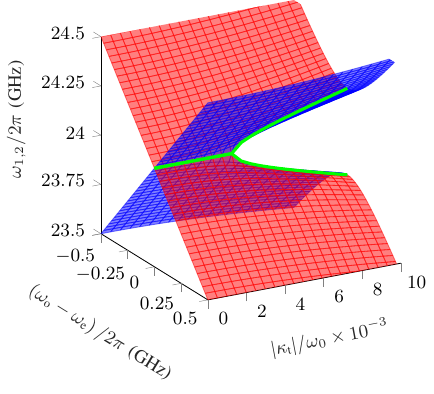}\label{fig:Riemann_Re}}
    \subfigure[]{\includegraphics{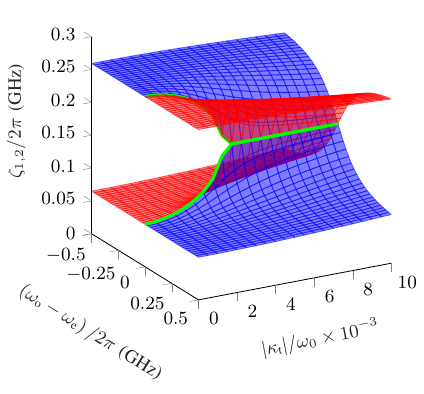}\label{fig:Riemann_Im}}

    \caption{Riemann surfaces corresponding to the complex coupled temporal eigenfrequencies in Eq.~\eqref{eq:xi12} versus $(\omega_\odd-\omega_\even)/2\pi$ and $\kappa_\temp/\omega_0$ with $\xi_\even/2\pi = (24 + j0.028)~\SI{}{GHz}$ and $\xi_\odd/2\pi = (24+j0.287)~\SI{}{GHz}$ for the SFP-P-LWA in \figref{fig:P-LWA}.}
    \label{fig:RiemannSheets}
\end{figure}

\pati{Complex Coupled Eigenfrequencies}
The complex coupled eigenfrequencies may then be obtained by substituting Eq.~\eqref{eq:cpl_mode_psi} and Eqs.~\eqref{eq:kappa_oe} into Eq.~\eqref{eq:cme}. This results in the eigenvalue system
\begin{subequations}\label{eq:P-LWA_matr_syst}
    \begin{equation}
        \begin{bmatrix} 
            \xi_\even & \kappa_\temp\\
            \kappa_\temp^* & \xi_\odd
        \end{bmatrix}
        \begin{bmatrix} 
            \psi_{\temp,1} \\ \psi_{\temp,2}
        \end{bmatrix}
        =
        \xi
        \begin{bmatrix} 
            \psi_{\temp,1} \\ \psi_{\temp,2}
        \end{bmatrix},
    \end{equation}
whose eigenvalues are found to be
\begin{align}\label{eq:xi12}
 \xi_{1,2} = \dfrac{\xi_\odd + \xi_\even}{2} \pm \sqrt{\kappa_\temp^2 + \left(\dfrac{\xi_\odd-\xi_\even}{2}\right)^2}.
\end{align}
\end{subequations}
As expected, the system obtained in Eqs.~\eqref{eq:P-LWA_matr_syst} corresponds to the system described in Sec.~\ref{sec:PT}~\cite{Bender_PT_Book_2019} and is, hence, \emph{$\mathcal{PT}$-symmetric}; the formal proof is given in Appendix~\ref{app:PTS_all}, where the loss-loss eigensystem~\eqref{eq:P-LWA_matr_syst} is first transformed into a gain-loss system by a gauge transformation (Appendix~\ref{app:Passive_PT}) and where this new system is then shown to satisfy the condition~\eqref{eq:PTS_def} (Appendix~\ref{app:P-LWA_PT}). The eigenfrequencies in Eq.~\eqref{eq:xi12} may be separated in terms of real and imaginary parts, which, from Eq.~\eqref{eq:cpl_mode_xi}, correspond to the functions $\omega_{1,2}$ and $\zeta_{1,2}$, of $\kappa_\temp$, $\omega_\even$, $\omega_\odd$, $\zeta_\even$ and $\zeta_\odd$, each of which is a two-sheet Riemann surface with the upper and lower sheets corresponding to the positive and negative signs in the equations. The two surfaces are plotted versus $\omega_\odd-\omega_\even$ and $\kappa_\temp$ in \figref{fig:RiemannSheets} for the SFP structure in \figref{fig:P-LWA}.

\pati{Frequency Balancing and Exceptional Point}
Under the \emph{frequency balancing} condition,
\begin{equation}\label{eq:fbal}
    \omega_\odd = \omega_\even = \omega_0,
\end{equation}
which is known as a necessary condition to avoid a gap that would otherwise separate the backward and forward regimes of P-LWA radiation~\cite{Caloz_MetaTra_2006}, Eq.~\eqref{eq:xi12} reduces to
\begin{align}\label{eq:xi12_bal} 
    \xi_{1,2} = \omega_0 + j\zeta_\Sigma \pm \sqrt{\kappa_\temp^2 - \zeta_\Delta^2},
\end{align}
where 
\begin{subequations}\label{eq:zeta_PM}
\begin{align}
    \begin{split}\label{zeta_P}
        \zeta_\Sigma=(\zeta_\even+\zeta_\odd)/2,
    \end{split}\\
    \begin{split}\label{zeta_M}
        \zeta_\Delta=(\zeta_\even-\zeta_\odd)/2.
    \end{split}
\end{align}
\end{subequations}
This corresponds to the cut planes $\omega_\odd-\omega_\even=0$ that are highlighted by the green curves in \figref{fig:RiemannSheets}.\footnote{The whole system governed by Eqs.~\eqref{eq:P-LWA_matr_syst} is called $\mathcal{PT}$-symmetric by abuse of language. In reality, only the ``$\mathcal{PT}$-tuned''~\cite{El-Ganainy_NPHYS_01_2018} pitchfork curves corresponding to $\omega_\odd=\omega_\even$ and highlighted in green in the figure are $\mathcal{PT}$-symmetric according to the definition in Eq.~\eqref{eq:PTS_def}, as shown in Appendix~\ref{app:P-LWA_PT}.} These curves qualitatively correspond to the pitchfork spectrum in \figref{fig:PT_gl} and are therefore associated with the EP shown in \figref{fig:PT_gg} or \figref{fig:PT_gl}. That point is in fact the branch point of the Riemann surface. It is found by eliminating the radicand in \eqref{eq:xi12_bal}. This leads to the coupling coefficient
\begin{align}\label{eq:epc}
 \kappa_{\temp,\ep} = \zeta_\Delta = \dfrac{\zeta_\even-\zeta_\odd}{2},
\end{align}
whose insertion into Eq.~\eqref{eq:xi12_bal} reduces the eigenfrequency to
\begin{align}\label{eq:xi_EP} 
 \xi_{\ep} = \omega_0 + j\zeta_\Sigma.
\end{align}

\pati{$Q$-Balancing Condition}
Let us finally determine the coupled quality factors at the EP. The quality factor is generally defined as
\begin{align}\label{eq:Q12}
    Q_{1,2}=\frac{\omega_{\temp,1,2}}{2\zeta_{\temp,1,2}}.
\end{align}
Inserting Eq.~\eqref{eq:xi_EP} into this relation yields
\begin{align}\label{eq:Q_EP}
	Q_\ep = Q_1 = Q_2 = \dfrac{\omega_0}{2\zeta_\Sigma}= \dfrac{\omega_0}{\zeta_\even + \zeta_\odd} = \dfrac{2}{\dfrac{1}{Q_\odd}+\dfrac{1}{Q_\even}}.
    \end{align}
That is exactly the \emph{$Q$-balancing} condition that was found in~\cite{Otto_TAP_10_2014}, via circuit modeling, to be the second condition for solving the broadside issue!

\begin{figure}
    % \centering
    \setlength\figureheight{.32\textwidth}
    \setlength\figurewidth{.32\textwidth}
		
    \hspace{0.8cm}
    % \tikzsetnextfilename{Xi_CMT_SFP}
    % \subfigure[]{\input{figs/Xi_CMT_SFP.tikz}\label{fig:SFP_eigenF}}
    % \hspace{0.2cm}
    % \tikzsetnextfilename{Q_factor_CMT_SFP}
    % \subfigure[]{\input{figs/Q_factor_CMT_SFP.tikz}\label{fig:SFP_Q}}
		
   \subfigure[]{\includegraphics{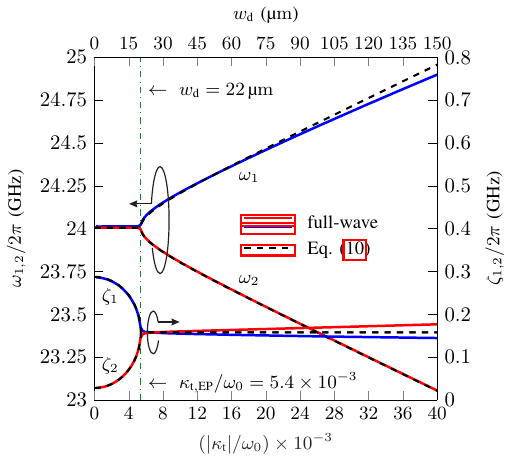}\label{fig:SFP_eigenF}}
   \subfigure[\hspace*{-0.8cm}]{\hspace{0.4cm}\includegraphics{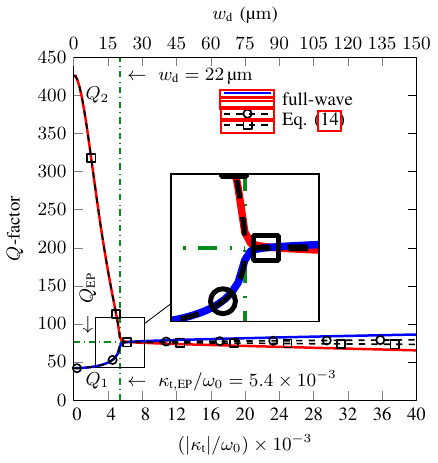}\label{fig:SFP_Q}}
    \caption{Analytical and full-wave (CST) results for the SFP P-LWA in \figref{fig:P-LWA}. (a)~Coupled complex eigenfrequencies [Eq.~\eqref{eq:xi12_bal}]. (b)~Coupled $Q$-factors [Eq.~\eqref{eq:Q12}].}
    \label{fig:SFP_results}
\end{figure}

\pati{Two-Dimensional Graphs}
Figure~\ref{fig:SFP_results} presents two-dimensional graphs for the complex eigenfrequencies (2D version of \figref{fig:RiemannSheets} in the plane \mbox{$(\omega_\odd - \omega_\even) = 0$} and for the quality factors using the temporal parameters in  Tab.~\ref{tab:temp_para}. The complex eigenfrequencies [Eq.~\eqref{eq:xi12_bal}], plotted in \figref{fig:SFP_eigenF}, exhibit exactly the loss-loss pitchfork spectrum in \figref{fig:PT_gl},\footnote{There is in fact a small difference: the slight divergence of $\zeta_1$ and $\zeta_2$ form $\zeta_\Sigma$ for increasing $\kappa_\spat$ in the broken $\mathcal{PT}$-symmetry region (right of the EP). This divergence is due to small detuning from $\omega_0$ as the level of asymmetry increases and is nonessential here insofar as it does not concern the EP.} with the EP at the junctions of the two pitchforks, while the $Q$-factors [Eq.~\eqref{eq:Q12}], plotted in \figref{fig:SFP_Q} merge at the EP.
\begin{table}[ht!]
	\centering
		\caption{Temporal frequency parameters (in GHz) obtained by full-wave simulation for the SFP P-LWA in \figref{fig:P-LWA}.}
		\begin{tabular}{c | c | c | c | c}
			$\omega_0/2\pi$ & $\xi_\even/2\pi$ & $\xi_\odd/2\pi$ & $\zeta_\Sigma/2\pi$ & $\zeta_\Delta/2\pi$ \\ \hline
			$24$	    	& $24 + j0.028$	   & $24 + j0.287$   & $0.1575$            & $-0.1295$
		\end{tabular}
	\label{tab:temp_para}
\end{table}

\pati{Summary}
In summary, we have demonstrated in this section, using coupled-mode theory, that a P-LWA is a \emph{$\mathcal{PT}$-symmetric system with an EP} and that this exceptional point corresponds to the satisfaction of the twofold \emph{frequency balancing and $Q$-balancing condition} found by a circuit approach in previous research as being the solution to the broadside issue~\cite{Otto_TAP_10_2011, Otto_TAP_04_2014, Otto_TAP_10_2014}. This is comforting, but not sufficient: our intention here is to establish a theory that is self-consistent.

% %%%%%%%%%%%%%%%%%%%%%%%%%%%%%%%%%%%%%
% %%%%% Temporal-Spatial Mapping  %%%%%
% %%%%%%%%%%%%%%%%%%%%%%%%%%%%%%%%%%%%%

\section{Time to Space Mapping}\label{sec:TS_mapping}
\pati{Motivation}
The temporal-frequency analysis performed in the previous section revealed that a P-LWA is a $\mathcal{PT}$-symmetric system and identified its EP. However, the related resonator or standing-wave regime does \emph{not} correspond to the operation regime of the antenna, which functions as a propagation or traveling-wave system. Therefore, we need to perform a temporal-frequency to spatial-frequency mapping in order to gain information on the physics of the EP and find out whether it indeed corresponds to the resolution of the broadside issue, as suggested from comparison with~\cite{Otto_TAP_10_2011, Otto_TAP_04_2014, Otto_TAP_10_2014} in the previous section. 

\pati{Procedure}
Such a mapping from temporal frequencies, $\xi=\omega+j\zeta$, to spatial frequencies,  $\gamma=\alpha+j\beta$, has been a subject of investigation for over a decade~\cite{Otto_MWSYM_06_2012,Dyab_ISAP_11_2015,Rahmeier_TAP_08_2021,Dehmollaian_TAP_10_2021}. This research culminated in a theory and procedure presented in~\cite{Dehmollaian_TAP_10_2021}, which leverages analytical continuity for mapping the $\xi$ and $\zeta$ complex planes to each other for arbitrary problems. The procedure is particularly simple when the solution of the system is available in analytical form in the starting plane, as is the case in our P-LWA analysis, in which case it reduces to the following two steps:
\begin{enumerate}
    \item replace the (complex) frequency variable $\xi$ by the (real) frequency variable $\omega$;
    \item replace the (real) variable $\beta\left(\xi\right)$ by $-j\gamma\left(\omega\right)$;
    \item solve the resulting equation for $\gamma$.
\end{enumerate}

\pati{$\kappa-\beta$ Relationship}
The mapping procedure must be applied here to the complex temporal frequency $\xi$. However, the currently available formula, given by Eq.~\eqref{eq:xi12_bal}, is not a function of $\beta$. This is because it has been derived for the $\Gamma$ point of the reciprocal space, $\Phi=\beta{p}=0$, corresponding to the broadside point of the P-LWA. Moving now to the complex spatial frequency ($\gamma=\alpha+j\beta$) plane to investigate the frequency scanning operation of the P-LWA naturally requires introducing the $\beta$ variable in Eq.~\eqref{eq:xi12_bal}. This can be done at this point by realizing that $\Phi=\beta p\neq{0}$ must act a contribution to coupling, viz., as part of $\kappa_\temp$ (in rad/s) in Eqs.~\eqref{eq:cme}, so that the total coupling coefficient decomposes as
\begin{align}\label{eq:kappa_beta}
    \kappa_\temp
    = \sqrt{\left(\dfrac{\omega_0\Phi}{2\pi}\right)^2 + \kappa_{\temp,\ga}^2} = \sqrt{\left(\dfrac{\omega_0\beta p}{2\pi}\right)^2 + \kappa_{\temp,\ga}^2},
\end{align}
where $\kappa_{\temp,\ga}$ is the coupling coefficient due to geometrical asymmetry [\figref{fig:rad_pat_bs}], while $\omega_0\Phi/2\pi=\omega_0\beta{p}/2\pi$ appears to be a contribution to the overall coupling that is due to propagation phase shifting across a unit cell. Note that this contribution disappears at broadside, where only geometrical asymmetry remains as the expected remedy to the broadside issue. 

\pati{Mapping}
Substituting Eq.~\eqref{eq:kappa_beta} into Eq.~\eqref{eq:xi12_bal}, we get
\begin{align}\label{eq:xi12_beta} 
 \xi_{1,2} = \omega_0 + j\zeta_\Sigma \pm \sqrt{\left(\dfrac{\omega_0\beta p}{2\pi}\right)^2 + \kappa_{\temp,\ga}^2 - \zeta_\Delta^2},
\end{align}
which, including now $\beta$, is an adequate formula for mapping. So, we proceed to the mapping procedure. First, we substitute $\xi_{1,2}$ by $\omega$ and $\beta$ by $-j\gamma$, and then we solve the resulting equation for $\gamma{p}$, which yields
\begin{align}\label{eq:gamma_t}
    \gamma_{1,2} p = \pm\dfrac{j2\pi}{\omega_0}\sqrt{\left(\omega - \omega_0 - j\zeta_\Sigma\right)^2 + \zeta_\Delta^2 - \kappa_{\temp,\ga}^2}.
\end{align}

\pati{Spatial Frequency at the Exceptional Point}
Interestingly, inserting the EP coupling factor at broadside, i.e., Eq.~\eqref{eq:epc} with $\kappa_{\temp,\ga}=\kappa_{\temp,\ep}$ into Eq.~\eqref{eq:gamma_t} reduces the dispersion relation to
\begin{subequations}\label{eq:gamma_12_p}
    \begin{align}\label{eq:gamma_t_ep}
        \gamma_{1,2;\ep} p = \pm\dfrac{j2\pi}{\omega_0}\sqrt{\left(\omega - \omega_0 - j\zeta_\Sigma\right)^2 },
    \end{align}
or
    \begin{align}\label{eq:gamma_ep}
        \gamma_{1,2;\ep} p = (\alpha + j\beta) p = \pm\dfrac{j2\pi}{\omega_0}\left(\omega - \omega_0 -j\zeta_\Sigma\right),
    \end{align}
\end{subequations}
so that
\begin{subequations}
    \begin{align}\label{eq:alpha_beta_ep}
        \beta_\ep p = \pm2\pi\dfrac{\omega-\omega_0}{\omega_0},
    \end{align}
    which retrieves the expected result $\beta(\omega=\omega_0)=0$, and
    \begin{align}
        \alpha_\ep p = \pm 2\pi \dfrac{\zeta_\Sigma}{\omega_0},
    \end{align}
    which is frequency-independent.
\end{subequations}
The frequency independence of $\alpha$, or flat $\alpha(\omega)$ curve, is a known condition for smooth transition across broadside~\cite{Paulotto_TAP_07_2009, Otto_TAP_10_2014}. However, this is still not a sufficient condition: given the asymmetry of the P-LWA structure [see \figref{fig:rad_pat_bs2}], which will be discussed in the next section, the antenna must exhibit different impedances at its two ports, and therefore the two ports cannot be matched to a single external impedance, $R_0$. So, the additional impedance matching condition will be satisfied at best at only one of the two ports.

\begin{figure}
    \centering
    \setlength\figureheight{.22\textwidth}
    \setlength\figurewidth{.18\textwidth}
    
    %\mbox{%
    %\hspace{0.8cm}
    %\tikzsetnextfilename{GammaT_Sym}
    %\subfigure[]{\input{figs/GammaT_Sym2.tikz}\label{fig:gamma_t0}}
    %\hspace{0.2cm}
    %\tikzsetnextfilename{GammaT_Asym_EP}
    %\subfigure[]{\input{figs/GammaT_Asym_EP2.tikz}\label{fig:gamma_topt}}
    %}
	\mbox{%
    \subfigure[]{\includegraphics{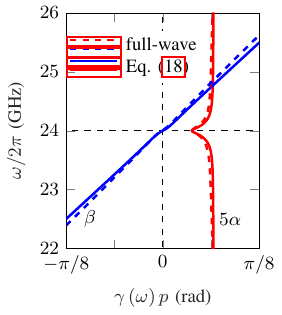}\label{fig:gamma_t0}}
    \hspace{-.2cm}
    \subfigure[]{\includegraphics{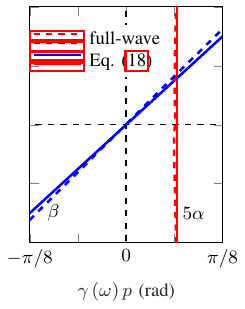}\label{fig:gamma_topt}}
    }
    \caption{Full-wave and analytically mapped complex spatial eigenfrequencies [Eq.~\eqref{eq:gamma_t}] for the SFP LWA in \figref{fig:P-LWA}. (a)~Symmetrical SFP LWA with $\kappa=0$ or $w_\textrm{d} = 0$. (b)~Asymmetrical SFP LWA with $\kappa_{\temp}=\kappa_{\temp,\ep}$ or $w_\textrm{d} = \SI{22}{\micro m}$ (see \figref{fig:SFP_results}).}
    \label{fig:gamma_t}
\end{figure}

\pati{Dispersion Relation Plotting}
Figure~\ref{fig:gamma_t} compares the theoretical\footnote{Remarkably, the theoretical curves in the figure can be plotted analytically, using Eq.~\eqref{eq:gamma_t}, \emph{on the basis of a single full-wave eigenfrequency simulation}, specifically the simulation of the initial, symmetric version of the P-LWA structure, which provides the even and odd eigenfrequencies. Indeed, the parameters $\omega_0$ and $\zeta_{\Sigma,\Delta}$ depend solely on $\omega_{\even,\odd}$ and $\zeta_{\even,\odd}$, according to Eqs.~\eqref{eq:fbal} and~\eqref{eq:zeta_PM}, respectively, while $\kappa_{\temp,\ga}$ is zero in the symmetric case and also depending only on $\zeta_{\even,\odd}$ at the EP, according to Eq.~\eqref{eq:epc}.} and full-wave simulated complex spatial eigenfrequency, $\gamma=\alpha+j\beta$. Figure~\ref{fig:gamma_t0} shows the manifestation of the broadside issue for a symmetrical P-LWA ($\kappa_t=0$) in terms of the deviation of $\alpha$ from a straight line, while \figref{fig:gamma_topt} shows the spatial spectrum of a P-LWA with optimal asymmetry in terms of a frequency-independent $\alpha$, consistent with Eq.~\eqref{eq:alpha_beta_ep}.

% %%%%%%%%%%%%%%%%%%%%%%%%%%%%%%%%%%%%%
% %%%% Spatial Analysis  %%%%
% %%%%%%%%%%%%%%%%%%%%%%%%%%%%%%%%%%%%%
\section{Space Analysis (Waveguide Problem)}\label{sec:SA_WP}

\begin{figure}
    \centering
    \subfigure[]{\includegraphics[]{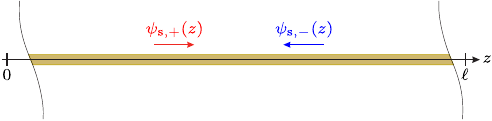}\label{fig:SCMT_Uniform}}
    \subfigure[]{\includegraphics[]{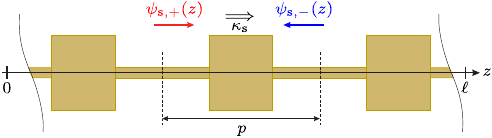}\label{fig:SCMT_Periodic}}
    \subfigure[]{\includegraphics[]{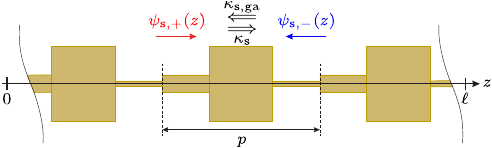}\label{fig:SCMT_Asym}}
    \caption{Three successive steps to determine the impedance of a P-LWA at its two ends. (a)~Uniform transmission line. (b)~Periodically loaded transmission line with symmetric unit cell. (c)~Same as (b) but with an asymmetric unit cell, as in \figref{fig:rad_pat_bs2}.}
    \label{fig:SCMT}
\end{figure}

\pati{Strategy of the Section}
We shall derive here formulas for the impedances at the two ends of an asymmetric P-LWA [see \figref{fig:rad_pat_bs2}] in the three successive and progressive steps indicated in \figref{fig:SCMT}, in order to find out whether the EP complex spatial frequency in Eq.~\eqref{eq:gamma_12_p} indeed corresponds to the resolution of the broadside issue at one of the two ends of the structure.

\pati{Uniform Transmission Line}
Let us start with a uniform transmission line, shown in \figref{fig:SCMT_Uniform}, which we assume to be dispersionless\footnote{The microwave transmission lines that constitute the backbone of typical P-LWAs, such as microstrip and coplanar transmission lines, are indeed essentially dispersionless in the frequency range of interest.} and matched. Such a structure admits \emph{forward and backward waves}, with complex wavenumbers $+\gamma_0$ and $-\gamma_0$, respectively. The corresponding waveforms are\footnote{The s subscript in the following expressions emphasizes the related quantities pertains to the domain of spatial complex frequencies, or waveguide regime.}
\begin{subequations}
    \begin{align}\label{eq:waveform_SF}
        \psi_{\spat;\pm}\left(t,z\right) = e^{j\omega t}e^{\mp\gamma_0z},
    \end{align}
    where
    \begin{equation}
        \gamma_0=\alpha+j\beta.
    \end{equation}
\end{subequations}
Given the uniformity of the transmission line and matching assumption, these waves are not coupled and satisfy then the uncoupled equations
\begin{subequations}\label{eq:scmt}
 \begin{align}   
  \begin{split}     
    \dfrac{d{\psi_{\spat,+}}}{dz} = -\gamma_0 \psi_{\spat,+},
  \end{split}\\
  \begin{split}    
    \dfrac{d{\psi_{\spat,-}}}{dz} = \gamma_0\psi_{\spat,-},
  \end{split} 
 \end{align}
\end{subequations}
which are the spatial counterparts of Eqs.~\eqref{eq:tcmt}.

\pati{Periodic Perturbations (Grating)}
Let us move on to the symmetric periodic structure in \figref{fig:SCMT_Periodic}, which represents the P-LWA category in \figref{fig:rad_pat_bs1}. Now the forward and backward waves are coupled by periodic scattering along the structure and become therefore an infinite set of \emph{forward and backward space harmonics}. Consequently, Eqs.~\eqref{eq:scmt} are augmented to the coupled equations~\cite{Orfanidis_ElectroWaveAnt_Book_2016},
\begin{subequations}\label{eq:sczf}
    \begin{align}   
        \dfrac{d{\psi_{\spat,1}}}{dz} = -\gamma_0\psi_{\spat,1}-j\kappa_\spat\exp\left(-j\frac{4\pi}{p}z\right)\psi_{\spat,2},
    \end{align}
    \begin{align}    
        \dfrac{d{\psi_{\spat,2}}}{dz} = j\kappa_\spat^*\exp\left(-j\frac{4\pi}{p}z\right)\psi_{\spat,1}+\gamma_0\psi_{\spat,2},
    \end{align}
\end{subequations}
where the coupling terms include the periodicity of the structure, consistent with the fact that coupling is periodic. In order to access the eigenvalues, we may apply the transformation~\cite{Orfanidis_ElectroWaveAnt_Book_2016}
\begin{align}\label{eq:scmt_transformation}
    \begin{bmatrix}
        \psi_{\spat,1} \\
        \psi_{\spat,2}
    \end{bmatrix} 
    = j 
    \begin{bmatrix}
         \exp\left(-j2\pi/p z\right) & 0\\
        0 & \exp\left(j2\pi/p z\right)
    \end{bmatrix}
    \begin{bmatrix}
        \Psi_{\spat,1}\\
        \Psi_{\spat,2}
    \end{bmatrix} 
\end{align}
to Eqs.~\eqref{eq:sczf}, which allows us to factor out the term $\exp\left(j4\pi/p z\right)$ and leads to the new coupled system
\begin{subequations}\label{eq:scmt_per_transformed}
    \begin{equation}
        \dfrac{d}{dz}
        \begin{bmatrix}
            \Psi_{\spat,1}\\
            \Psi_{\spat,2}
        \end{bmatrix} 
        = 
        \begin{bmatrix}
            -\gamma_{0,\textrm{p}}& -j\kappa_\spat\\
            j\kappa_\spat^* & \gamma_{0,\textrm{p}}
        \end{bmatrix}
        \begin{bmatrix}
            \Psi_{\spat,1}\\
            \Psi_{\spat,2}
        \end{bmatrix},
    \end{equation}
    where
    \begin{align}
        \gamma_{0,\textrm{p}} = \alpha + j\left(\beta - \dfrac{2\pi}{p}\right),
    \end{align}
\end{subequations}
and whose eigensolutions are found to be
\begin{align}\label{eq:gamma_per}
    \gamma_{1,2}\left(\omega\right) = \pm\sqrt{\gamma_{0,\textrm{p}}^2 - |\kappa_\spat|^2}.
\end{align}

\pati{Asymmetric Perturbations}
Let us finally consider the asymmetric periodic structure in \figref{fig:SCMT_Asym}, which represents the P-LWA category in \figref{fig:rad_pat_bs2}. The related coupled equations are formally identical to those of Eqs.~\eqref{eq:scmt_per_transformed}, except for an additional coupling term accounting for coupling due to the geometrical asymmetry, $\kappa_{\spat,\ga}$. Thus, the coupled mode system reads
\begin{align}\label{eq:scmt_asym}
    \begin{split}
        \dfrac{d}{dz}
        \begin{bmatrix}
            \Psi_{\spat,1}\\
            \Psi_{\spat,2}
        \end{bmatrix} 
        = & 
        \begin{bmatrix}
            -\gamma_{0,\textrm{p}} & -j(\kappa_\spat+j\kappa_{\spat,\ga})\\
            j(\kappa_\spat^*-j\kappa_{\spat, \ga}^*) & \gamma_{0,\textrm{p}}
        \end{bmatrix}
        \begin{bmatrix}
            \Psi_{\spat,1}\\
            \Psi_{\spat,2}
        \end{bmatrix}
    \end{split}
\end{align}
and the related eigenvalues are found to be
\begin{align}\label{eq:gamma_asym}
    \gamma_{1,2}\left(\omega\right) = \pm\sqrt{\gamma_{0,\textrm{p}}^2 - |\kappa_\spat|^2 + |\kappa_{\spat,\ga}|^2}.
\end{align}

\pati{Spatial-Temporal Parameters Relationship}
Comparing the temporal [Eq.~\eqref{eq:gamma_t}] and spatial frequencies [Eq.~\eqref{eq:gamma_asym}] yields
\begin{subequations}\label{eq:spat_temp_para}
\begin{equation}\label{eq:gamma_ts}
    \begin{split}      
    \gamma_{0,\textrm{p}}^2 &- |\kappa_\spat|^2 + |\kappa_{\spat,\ga}|^2  \\
    & = -\left(\dfrac{2\pi}{\omega_0 p}\right)^2\left[\left(\omega - \omega_0 - j\zeta_\Sigma\right)^2 + \zeta_\Delta^2 - \kappa_{\temp,\ga}^2\right],
    \end{split}
\end{equation}
or
\begin{align}
    \begin{split}\label{eq:gamma_sym_zeta}
    \gamma_{0,\textrm{p}} &= j2\pi\frac{\omega - \omega_0 - j\zeta_\Sigma}{\omega_0 p},
    \end{split}\\
    \begin{split}\label{eq:kappa_spat_zeta}
    |\kappa_\spat|        &= \frac{2\pi\zeta_\Delta}{\omega_0 p},
    \end{split}\\
    \begin{split}\label{eq:kappa_spat_EP_zeta}
    |\kappa_{\spat,\ga}|  &= \frac{2\pi\kappa_{\temp,\ga}}{\omega_0 p},
    \end{split}
\end{align}
and hence
\begin{align}
    |\kappa_{\spat,\ep}| = \frac{2\pi\kappa_{\temp,\ep}}{\omega_0 p}.
\end{align}
\end{subequations}
These relations are useful because they directly provide the spatial quantities in Sec.~\ref{sec:SA_WP} from the temporal quantities in Sec.~\ref{sec:TA_RP} without requiring the space to time mapping in Sec.~\ref{sec:TS_mapping}.

\begin{figure}
    \centering
    \includegraphics[]{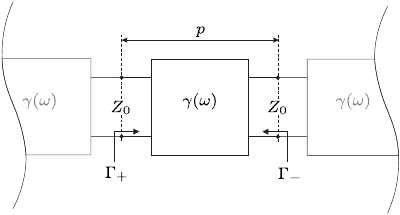}
    \caption{Equivalent two-port network model for the calculation of the forward and backward Bloch impedances, $Z_\bloch^\pm$, in Eq.~\eqref{eq:ZB} with $Z_0$ being the homogenized reference impedance.}
    \label{fig:TL_homogen}
\end{figure}

\pati{Impedance Derivation}
We may now derive the sought-after impedance formulas for the asymmetric P-LWA. The waves at the two ends of the structure, assumed to be of length $\ell$,\footnote{The total length of a periodic structure is $Np$, where $N$ is the total number of unit cells.} are related by the transfer matrix $\exp(jM_\spat\ell)$ as~\cite{Orfanidis_ElectroWaveAnt_Book_2016}
\begin{subequations}
\begin{align}
    \begin{bmatrix}
        \Psi_{\spat,1}\left(0\right)\\
        \Psi_{\spat,2}\left(0\right)
    \end{bmatrix} 
        = \exp\left(jM_\spat\ell\right)
    \begin{bmatrix}
        \Psi_{\spat,1}\left(\ell\right)\\
        \Psi_{\spat,2}\left(\ell\right)
    \end{bmatrix},
\end{align}
where
\begin{align}
    M_\spat
    = \begin{bmatrix}
    -\gamma_{0,\textrm{p}} & -j(\kappa_\spat + \kappa_{\spat, \ga})\\
    j(\kappa_\spat-\kappa_{\spat, \ga}) & \gamma_{0,\textrm{p}}
    \end{bmatrix},
\end{align}
with $(\kappa_\spat-\kappa_{\spat,\ga})^*=(\kappa_\spat-\kappa_{\spat,\ga})$ from reciprocity, and
\begin{align}\label{eq:transfereM}
    \begin{split}
    \exp\left(jM_\spat\ell\right)
    =
    \begin{bmatrix}
    U_{11} & U_{12}\\
    U_{21} & U_{22}
    \end{bmatrix}
    \end{split}
    = \nonumber\\
    \begin{split}
    \begin{bmatrix}
    \cosh\left(\gamma\ell\right)+j\dfrac{\gamma_{0,\textrm{p}}}{\gamma}\sinh\left(\gamma\ell\right) & j\dfrac{(\kappa_\spat + \kappa_{\spat, \ga})}{\gamma}\sinh\left(\gamma\ell\right)\\
    -j\dfrac{(\kappa_\spat - \kappa_{\spat, \ga})}{\gamma}\sinh\left(\gamma\ell\right) & \cosh\left(\gamma\ell\right)-j\dfrac{\gamma_{0,\textrm{p}}}{\gamma}\sinh\left(\gamma\ell\right)
    \end{bmatrix},
    \end{split}
\end{align}
\end{subequations}
where $\gamma$ was defined by Eq.~\eqref{eq:gamma_asym}. Rearranging the previous system in terms of scattering parameters yields
\begin{subequations}
    \begin{align}
        \begin{bmatrix}
             \Psi_{\spat,2}\left(0\right)\\
             \Psi_{\spat,1}\left(\ell\right)
        \end{bmatrix} 
        = 
        \begin{bmatrix}
            \Gamma^+ & T\\
            T & \Gamma^-
        \end{bmatrix}
        \begin{bmatrix}
             \Psi_{\spat,1}\left(0\right)\\
             \Psi_{\spat,2}\left(\ell\right)
        \end{bmatrix},
    \end{align}
\end{subequations}
from which we obtain the forward reflection coefficient
\begin{subequations}\label{eq:s_par}
    \begin{equation}
        \Gamma_+ = \dfrac{U_{21}}{U_{11}}=-\frac{j\dfrac{(\kappa_\spat - \kappa_{\spat, \ga})}{\gamma}\sinh\left(\gamma\ell\right)}{\cosh\left(\gamma\ell\right)+j\dfrac{\gamma_{0,\textrm{p}}}{\gamma}\sinh\left(\gamma\ell\right)},
    \end{equation}
    the backward reflection coefficient
    \begin{equation}
        \Gamma_- = -\dfrac{U_{12}}{U_{22}}=-\frac{j\dfrac{(\kappa_\spat + \kappa_{\spat, \ga})}{\gamma}\sinh\left(\gamma\ell\right)}{\cosh\left(\gamma\ell\right)-j\dfrac{\gamma_{0,\textrm{p}}}{\gamma}\sinh\left(\gamma\ell\right)},
    \end{equation}
    and the transmission coefficient
    \begin{equation}
        T = \dfrac{1}{U_{11}}=\frac{1}{\cosh\left(\gamma\ell\right)+j\dfrac{\gamma_{0,\textrm{p}}}{\gamma}\sinh\left(\gamma\ell\right)}.
    \end{equation}
\end{subequations}
Moreover, the Bloch impedance in the forward and backward directions, $Z_\bloch^\pm$, can be calculated after transforming the scattering matrix [Eq.~\eqref{eq:transfereM}] into the ABCD-matrix, as~\cite{Pozar_MicroEng_Book_2011}
\begin{align}\label{eq:ZB}
    \dfrac{Z_\bloch^\pm}{Z_0} = \dfrac{-2B}{A-D\mp\sqrt{\left(A+D\right)^2 - 4}},
\end{align}
where $Z_0$ is the reference impedance\footnote{The reference impedance $Z_0$ for the simple transmission line in \figref{fig:SCMT_Uniform} is the characteristic impedance, whereas for the periodic structures in Figs.~\ref{fig:SCMT_Periodic} and \ref{fig:SCMT_Asym}, $Z_0$ depends on the impedance step and the reference location.} for a two-port network model, as shown in \figref{fig:TL_homogen}.

\begin{figure}
    \centering	
    
	%\setlength\figureheight{.22\textwidth}
    %\setlength\figurewidth{.18\textwidth}
    %\tikzsetnextfilename{GammaS_ZB_sym}
    %\subfigure[]{\input{figs/GammaS_ZB_sym2.tikz}\label{fig:gamma_ZB_sym}}

    %\tikzsetnextfilename{GammaS_ZB_asym_EP}
    %\subfigure[]{\input{figs/GammaS_ZB_asym_EP2.tikz}\label{fig:gamma_ZB_EP}}
		
    \subfigure[]{\includegraphics[]{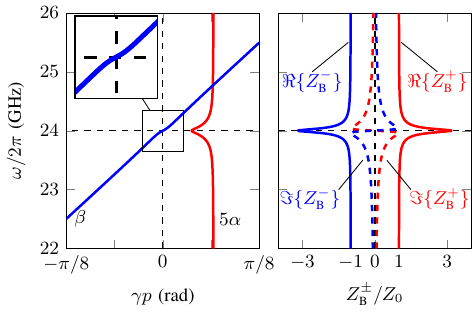}\label{fig:gamma_ZB_sym}}
    \subfigure[]{\includegraphics[]{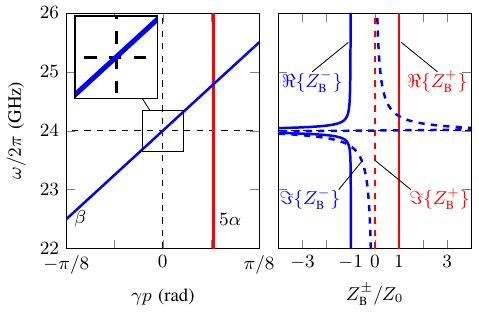}\label{fig:gamma_ZB_EP}}
    \caption{Analytical results for the spatial eigenfrequency [Eq.~\eqref{eq:gamma_asym}, using Tab.~\ref{tab:spat_para}] and Bloch impedance [Eq.~\eqref{eq:ZB}]. (a)~Symmetric structure with $\kappa_\spat = 0$ [\figref{fig:SCMT_Periodic}] and (b)~asymmetric structure at EP  with $\kappa_{\spat,\ga} = \kappa_{\spat,\ep} = \zeta_\Delta/\omega_0p$ [\figref{fig:SCMT_Asym}].}
    \label{fig:gamma_ZB_spat}
\end{figure}

\pati{Numerical Results}
Figure~\ref{fig:gamma_ZB_spat} plots the spatial eigenfrequency $\gamma$ and the normalized Bloch impedances $Z_\bloch^\pm/Z_0$ for the spatial parameters listed in Tab.~\ref{tab:spat_para}. In the symmetric case ($\kappa_\spat = 0$), corresponding to~\figref{fig:gamma_ZB_sym}, the forward and backward impedances\footnote{The negative sign in the backward impedance is simply due to the oppositely flowing ($-z$) current in the fixed coordinate system.} are, as expected, identical and peaking at $\omega_0/2\pi=\SI{24}{GHz}$. Meanwhile, the leakage factor ($\alpha$) shows a dip and the propagation constant ($\beta$) exhibits slight nonlinearity at $\omega_0$, both indicating a $Q$-mismatch. In contrast, in the EP-asymmetric case ($\kappa_{\spat,\ga} = \kappa_{\spat,\ep}$) [\figref{fig:gamma_ZB_EP}], the forward and backward Bloch impedances are different, as expected from the unequal forward and backward reflection coefficients ($\Gamma_+ \neq \Gamma_-$) [Eq.~\eqref{eq:s_par}]. While $Z_\bloch^-$ is highly dispersive, $Z_\bloch^+$ is \emph{frequency-independent}, consistently with
\begin{equation}
    \Gamma_+(\kappa_{\spat,\ga} = \kappa_{\spat,\ep})=0,
\end{equation} 
which reveals that the P-LWA is matched in the forward direction. Moreover, the leakage factor and propagation constant are perfectly flat and linear, respectively, confirming the expected resolution of the broadside issue, for the source placed at one of the two ends of the structure. This completes our $\mathcal{PT}$-symmetry-based demonstration that a frequency-balanced and $Q$-balanced P-LWA provides seamless scanning through broadside if fed at its ``good'' input.
\begin{table}[ht!]
	\centering
		\caption{Spatial parameters obtained from the temporal parameters in Tab.~\ref{tab:temp_para} using Eqs.~\eqref{eq:spat_temp_para}. All the units are in rad/m.}
		\begin{tabular}{c | c | c}
			$\gamma_{0,\textrm{p}}(\omega=\omega_0)$ & $|\kappa_\spat|$ & $|\kappa_{\spat,\ep}|$ \\ \hline
			$6.3072$	    	                     & $5.1816$	        & $5.1816$  
		\end{tabular}
	\label{tab:spat_para}
\end{table}
%%%%%%%%%%%%%%%%%%%%%%%%%%%%%%%%%%%%%
%%%%%%%%%%%% Conclusion  %%%%%%%%%%%%
%%%%%%%%%%%%%%%%%%%%%%%%%%%%%%%%%%%%%
\section{Conclusion}\label{sec:concl}
We have presented an electromagnetic theory of seamless-scanning P-LWAs based on the concept of $\mathcal{PT}$-symmetry. After demonstrating that a P-LWA is a $\mathcal{PT}$-symmetric system, we have shown that the long-standing issue of gain degradation at broadside is resolved by introducing in the P-LWA's unit cell an amount of geometrical asymmetry that corresponds to the EP of the $\mathcal{PT}$-symmetry spectrum.

This theory provides a solid foundation for the twofold condition of frequency balancing and $Q$-balancing that was discovered using circuit modeling in previous works. Moreover, it represents an original contribution to the field of $\mathcal{PT}$-symmetry by extending the concept from single-space to double-space, space-time $\mathcal{PT}$-symmetry.

%%%%%%%%%%%%%%%%%%%%%%%%%%%%%%%%%%%%%
%%%%%%%%%%%% Appendices  %%%%%%%%%%%%
%%%%%%%%%%%%%%%%%%%%%%%%%%%%%%%%%%%%%
\appendices

\section{Appendix: P-LWA \texorpdfstring{$\mathcal{PT}$}{PT}-Symmetry}\label{app:PTS_all}
\renewcommand\thesubsection{\thesection.\Roman{subsection}}
\renewcommand\thesubsectiondis{\thesection.\Roman{subsection}}

The $\mathcal{PT}$-Symmetry of a P-LWA may be shown in two steps: the transformation of its loss-loss eigenspectrum into a gain-loss eigenspectrum, which will be performed in Sec.~\ref{app:Passive_PT}, and the $\mathcal{PT}$-symmetry demonstration of that equivalent gain-loss system, which will be performed in Sec.~\ref{app:P-LWA_PT}.

\subsection{Gauge Transformation from Passive \\ to Active \texorpdfstring{$\mathcal{PT}$}{PT}-Symmetry}\label{app:Passive_PT}

Inserting Eqs.~\eqref{cplx_eo_freq} with balanced frequencies [Eqs.~\eqref{eq:fbal}: $\omega_\even=\omega_\odd=\omega_0$] and real coupling factor ($\kappa_\temp^*=\kappa_\temp$) into Eq.~\eqref{eq:cme} yields the Schr\"{o}dinger equation
\begin{subequations}
    \begin{align}\label{eq:tcmt_fbal}
        -j\dfrac{d}{dt}
        \begin{bmatrix}
            \psi_{\temp,1}\\
            \psi_{\temp,2}
        \end{bmatrix}
        =
            \mathcal{H}
        \begin{bmatrix}
            \psi_{\temp,1}\\
            \psi_{\temp,2}
        \end{bmatrix},
    \end{align}
    with the Hamiltonian
    \begin{align}\label{eq:H_ll}
        \mathcal{H}=
        \begin{bmatrix} 
            \omega_0+j\zeta_\even & \kappa_\temp\\
            \kappa_\temp & \omega_0 + j\zeta_\odd
        \end{bmatrix}.
    \end{align}
\end{subequations}
Then, inserting the gauge transformation\cite{Bender_PT_Book_2019}
\begin{align}\label{eq:gauge}
    \begin{bmatrix}
        \psi_{\temp,1}\\
        \psi_{\temp,2}
    \end{bmatrix}
    = \exp\left(\zeta_{\Sigma t}\right)
    \begin{bmatrix}
        \psi_{\temp,1}^\prime\\
        \psi_{\temp,2}^\prime
    \end{bmatrix},
\end{align}
where $\psi_{\temp,1}^\prime$ and $\psi_{\temp,2}^\prime$ are the gauge-transformed fields and $\zeta_\Sigma$ is given by Eq.~\eqref{zeta_P}, into Eq.~\eqref{eq:tcmt_fbal}, we get
\begin{align}\label{eq:gauged_H}
    \begin{split}
        j\dfrac{d}{dt}
        \begin{bmatrix}
            \psi_{\temp,1}^\prime\\
            \psi_{\temp,2}^\prime
        \end{bmatrix}
        &=
        \begin{bmatrix} 
            \omega_0+j\zeta_\even & \kappa_\temp\\
            \kappa_\temp & \omega_0 + j\zeta_\odd
        \end{bmatrix}
        \begin{bmatrix}
            \psi_{\temp,1}^\prime\\
            \psi_{\temp,2}^\prime
        \end{bmatrix}
        -j\zeta_\Sigma
        \begin{bmatrix}
            \psi_{\temp,1}^\prime\\
            \psi_{\temp,2}^\prime
        \end{bmatrix}
    \end{split}\nonumber\\
    \begin{split}
        &=
        \begin{bmatrix} 
             \omega_0+j\zeta_\Delta & \kappa_\temp\\
             \kappa_\temp & \omega_0-j\zeta_\Delta
        \end{bmatrix}
        \begin{bmatrix}
            \psi_{\temp,1}^\prime\\
            \psi_{\temp,2}^\prime
        \end{bmatrix},
    \end{split}
\end{align}
where $\zeta_\Sigma$ is given by Eq.~\eqref{zeta_M}. The matrix
\begin{equation}\label{eq:H_gl}
    \mathcal{H}'
    =\begin{bmatrix} 
        \omega_0+j\zeta_\Delta & \kappa_\temp\\
        \kappa_\temp & \omega_0-j\zeta_\Delta
    \end{bmatrix}
\end{equation}
in this relation is the Hamiltonian of the gauge-transformed version of $\mathcal{H}$ in Eq.~\eqref{eq:H_ll}. While $\mathcal{H}$ represents a $\mathcal{PT}$-symmetric loss-loss system, recognized by is two positive signs, $\mathcal{H}'$ represents a $\mathcal{PT}$-symmetric gain-loss system, recognized by its negative and positive signs.

\subsection{Demonstration of the P-LWA \texorpdfstring{$\mathcal{PT}$}{PT}-Symmetry}\label{app:P-LWA_PT}
A P-LWA antenna is governed by the loss-loss Hamiltonian $\mathcal{H}$ in Eq.~\eqref{eq:H_ll} or by its gauge-transformed version $\mathcal{H}'$ in Eq.~\eqref{eq:H_gl}. Let consider $\mathcal{H}'$. According to Eq.~\eqref{eq:PTS_def}, and noting that $\mathcal{P}^*=\mathcal{P}$, a system described by $\mathcal{H}'$ is $\mathcal{PT}$-symmetric if $\mathcal{P}\mathcal{H}'^*\mathcal{P}=\mathcal{H}'$. Let us test this condition:
\begin{align}\label{eq:PTH_commute}
    \begin{split}
        \mathcal{P}\mathcal{H^\prime}^*\mathcal{P}
        &=
        \begin{bmatrix} 
            0 & 1\\
            1 & 0
        \end{bmatrix} 
        \begin{bmatrix} 
             \omega_0+j\zeta_\Delta & \kappa_\temp\\
             \kappa_\temp & \omega_0 - j\zeta_\Delta
        \end{bmatrix}^*
        \begin{bmatrix} 
            0 & 1\\
            1 & 0
        \end{bmatrix}
    \end{split}\nonumber \\
    \begin{split}
       & =
       \begin{bmatrix} 
            0 & 1\\
            1 & 0
        \end{bmatrix} 
        \begin{bmatrix}
             \omega_0-j\zeta_\Delta & \kappa_\temp\\
             \kappa_\temp & \omega_0 + j\zeta_\Delta
        \end{bmatrix}
        \begin{bmatrix} 
            0 & 1\\
            1 & 0
        \end{bmatrix}
    \end{split}\nonumber \\
    \begin{split}
        &= 
        \begin{bmatrix} 
             \omega_0+j\zeta_\Delta & \kappa_\temp\\
             \kappa_\temp & \omega_0 - j\zeta_\Delta
        \end{bmatrix}
        = \mathcal{H^\prime}.
    \end{split}
\end{align}
The condition is satisfied. So, a frequency-balanced P-LWA is a $\mathcal{PT}$-symmetric system.

Note that $\mathcal{PT}$-symmetry occurs \emph{only in the balanced-frequency} regime. Indeed outside of this regime, Eqs.~\eqref{eq:PTH_commute} become
\begin{align}
    \begin{split}
        \mathcal{P}\mathcal{H^\prime}^*\mathcal{P}
        &=
        \begin{bmatrix} 
            0 & 1\\
            1 & 0
        \end{bmatrix} 
        \begin{bmatrix} 
             \omega_\even+j\zeta_\Delta & \kappa_\temp\\
             \kappa_\temp & \omega_\odd - j\zeta_\Delta
        \end{bmatrix}^*
        \begin{bmatrix} 
            0 & 1\\
            1 & 0
        \end{bmatrix}
    \end{split}\nonumber \\
    \begin{split}
       & =
       \begin{bmatrix} 
            0 & 1\\
            1 & 0
        \end{bmatrix} 
        \begin{bmatrix}
             \omega_\even-j\zeta_\Delta & \kappa_\temp\\
             \kappa_\temp & \omega_\odd + j\zeta_\Delta
        \end{bmatrix}
        \begin{bmatrix} 
            0 & 1\\
            1 & 0
        \end{bmatrix}
    \end{split}\nonumber \\
    \begin{split}
        &= 
        \begin{bmatrix} 
             \omega_\odd+j\zeta_\Delta & \kappa_\temp\\
             \kappa_\temp & \omega_\even - j\zeta_\Delta
        \end{bmatrix}
        \neq \mathcal{H^\prime}.
    \end{split}
\end{align}
This fact is also apparent in the Riemann sheets of \figref{fig:RiemannSheets} and consistent with the well-known necessity of frequency balancing for P-LWA seamless radiation through broadside~\cite{Caloz_MetaTra_2006}.

% \ifCLASSOPTIONcaptionsoff
%   \newpage
% \fi

% %%%%%%%%%%%%%%%%%%%%%%%%%%%%%%%%%%%%%%%%%%%%%%%%%%%%%%%%%%%%%%%%%%%%%%%%%%%%%%%%%%%%%%%
% %%%%%%%%%%%%%%%%%%%%%%%%%%      Bibliography     %%%%%%%%%%%%%%%%%%%%%%%%%%%%%%%%%%%%%%
% %%%%%%%%%%%%%%%%%%%%%%%%%%%%%%%%%%%%%%%%%%%%%%%%%%%%%%%%%%%%%%%%%%%%%%%%%%%%%%%%%%%%%%%
\bibliographystyle{IEEEtran}
\bibliography{PT-Sym_LWA}

\end{document}